\documentclass{NDSProceedings}
\usepackage{graphicx}
\def\ave#1{\left\langle{#1}\right\rangle}

\begin{document}

\ProcTitle{Unified Coupled-Channels and Hauser-Feshbach Model Calculation\\
           for Nuclear Data Evaluation}
\ProcAuthor{Toshihiko Kawano}
\ProcInstitute{Los Alamos National Laboratory, Los Alamos, NM 87545, USA}
\ProcEmail{kawano@lanl.gov}

\ProcAbstract{%
We present an overview of the coupled-channels optical model and the
Hauser-Feshbach theory code CoH$_3$, which focuses on the nuclear
reaction calculations in the keV to tens of MeV region with special
attention to the nuclear deformation. The code consists of three major
sections that undertake the one-body potential mean-field theory, the
coupled-channels optical model, and the Hauser-Feshbach statistical
decay. There are other complementary segments to perform the whole
nuclear reaction calculations, such as the direct/semidirect radiative
capture process, pre-equilibrium process, and prompt fission neutron
emission.}

\ProcMakeTitle

\section{Introduction}

Modern methodology for evaluating nuclear reaction data for medium to
heavy mass targets centers a statistical Hauser-Feshbach (HF) code in
the evaluation system. The HF theory with the width fluctuation
correction gives a compound nuclear reaction cross section when
resonances are strongly overlapped; in other words, an energy-averaged
cross section is calculated. The HF codes currently available in the
market, such as EMPIRE~\cite{INDC0603}, TALYS~\cite{Koning2008},
CCONE~\cite{Iwamoto2007}, and CoH$_3$~\cite{Kawano2010}, which are
capable for multi-particle evaporation from a compound nucleus,
provide complete information of nuclear reactions, not only the
reaction cross sections, but also the energy and angular distributions
of secondary particles, $\gamma$-ray production cross sections,
isomeric state productions, and so on. One of distinct features in
CoH$_3$ is a unique capability to combine the coupled-channels optical
model and the HF theory, where two methods are employed --- the
generalized transmission coefficients~\cite{Kawano2009} and the
Engelbrecht-Weidenm\"{u}ller transformation~\cite{Kawano2016}.
Recently a code comparison was performed amongst the developers of
EMPIRE, TALYS, CCONE, and CoH$_3$, which suggested that the inelastic
scattering cross section by CoH$_3$ tends to be slightly higher than
the other codes~\cite{Capote2017} due to this difference. This paper
outlines the reaction theories involved in CoH$_3$.

\section{CoH$_3$ Code Overview}

The CoH$_3$ code is written in C++, and it consists of about 200 source
files including 80 defined classes. For example, the simplest class is
{\tt ZAnumber} that has only two private member variables, the $Z$ and
$A$ numbers.  This class facilitates to calculate the $(Z,A)$ pair of
a compound nucleus emerging in a reaction chain, and it resembles the
traditional technique to represent the $(Z,A)$ pair by an index of
$1000Z + A$ in the FORTRAN77-age.

CoH$_3$ has its own optical model solver to generate the transmission
coefficients internally.  In the deformed nucleus case, a rotational
or vibrational model is employed for the coupled-channels (CC)
calculation. The nuclear structure properties are determined by
reading the nuclear structure database~\cite{RIPL3}.  At higher
excitation energies, we use the Gilbert-Cameron level density
formula~\cite{Gilbert1965} with updated parameters~\cite{Kawano2006}.
CoH$_3$ allows overlapping discrete levels inside the continuum
region. The width fluctuation correction is calculated by applying the
method of Moldauer~\cite{Moldauer1980} with the LANL updated
parameters~\cite{Kawano2014} based on GOE (Gaussian Orthogonal
Ensemble)~\cite{Kawano2015}. When strongly coupled channels exist, the
so-called Engelbrecht-Weidenm\"{u}ller transformation (EWT) is invoked
to diagonalize the $S$-matrix~\cite{Kawano2016}, and the width
fluctuation is calculated in the diagonalized channel (eigen-channel)
space.

Besides the main HF core part, the code consists of many models.  The
two-component exciton model~\cite{Kalbach1986,Koning2004} is used to
calculate the pre-equilibrium process. For fissioning nuclei, the
prompt fission neutron spectrum is calculated with the Madland-Nix
model~\cite{Madland1981} including pre-fission neutron emissions. The
direct/semidirect (DSD) neutron capture process is calculated with the
DSD model~\cite{Bonneau2007}.  There are three mean-field theories
included to calculate the single-particle wave-functions in a one-body
potential; FRDM (Finite Range Droplet
Model)~\cite{Moller1995,Moller2016}, HF-BCS (Hartree-Fock
BCS)~\cite{Bonneau2007}, and a simple spherical Woods-Saxon.

\begin{figure}[h]
  \begin{center}
    \begin{tabular}{cc}
      \resizebox{0.45\textwidth}{!}{\includegraphics{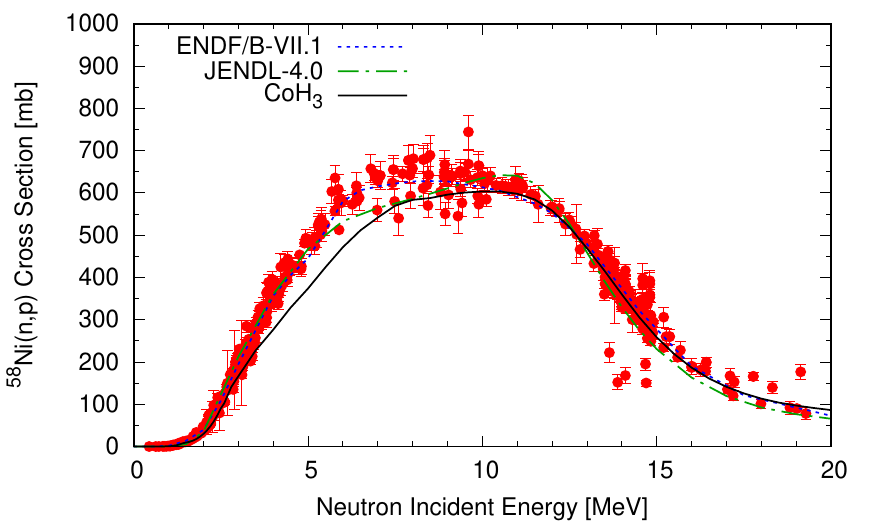}} &
      \resizebox{0.45\textwidth}{!}{\includegraphics{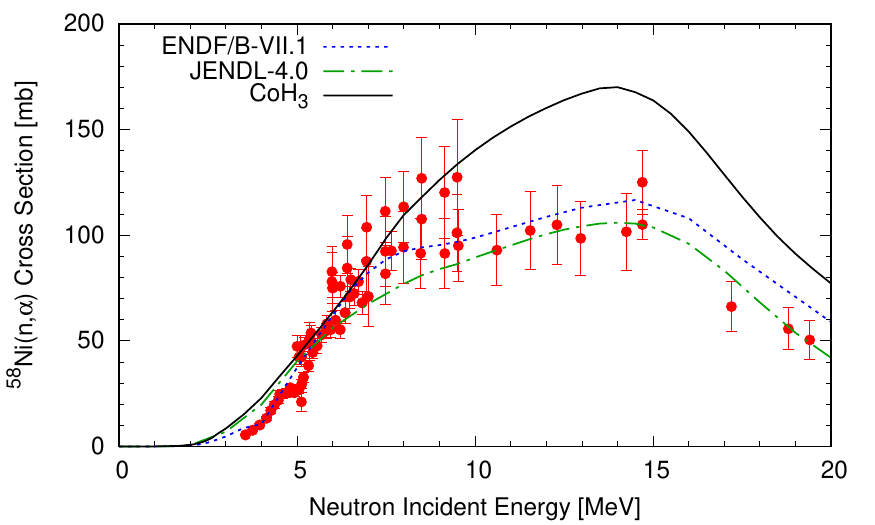}}\\
      \resizebox{0.45\textwidth}{!}{\includegraphics{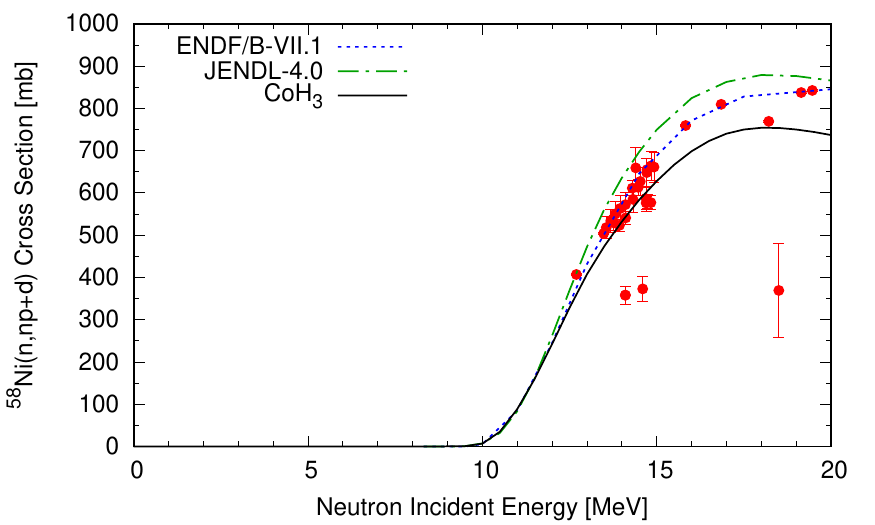}} &
      \resizebox{0.45\textwidth}{!}{\includegraphics{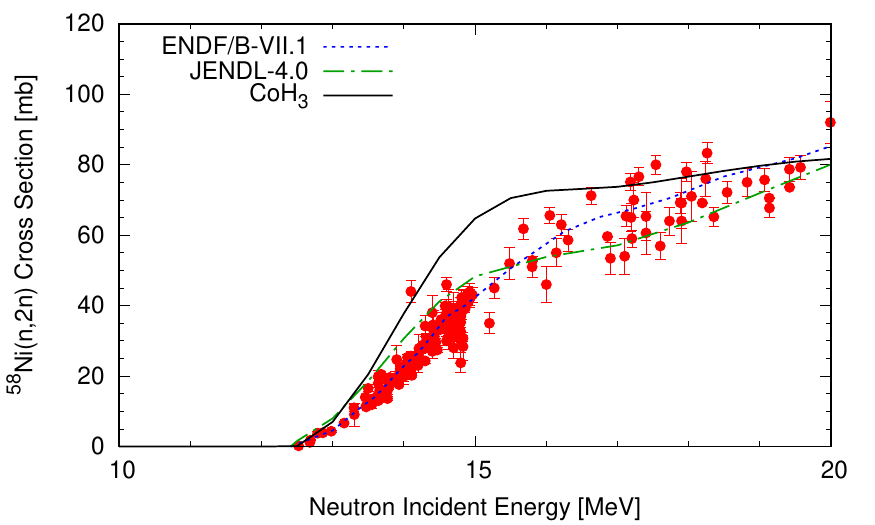}}\\
    \end{tabular}
  \end{center}
  \caption{CoH$_3$ default calculations for the neutron-induced reactions
    on $^{58}$Ni; (n,p), (n,$\alpha$), (n,np), and (n,2n) reactions. The (n,np)
    cross section includes the (n,d) reaction too.}
  \label{fig:Ni58}
\end{figure}

\begin{figure}
  \begin{center}
    \begin{tabular}{ccc}
      \resizebox{0.3\textwidth}{!}{\includegraphics{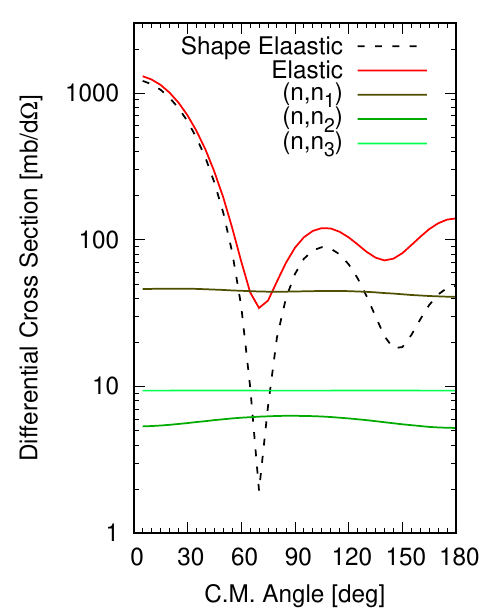}} &
      \resizebox{0.3\textwidth}{!}{\includegraphics{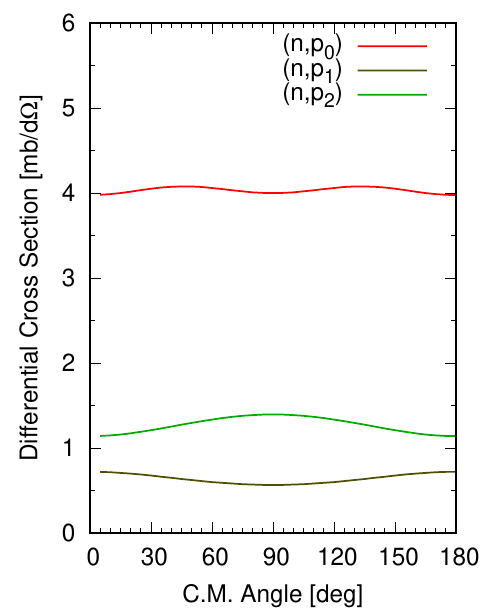}} &
      \resizebox{0.3\textwidth}{!}{\includegraphics{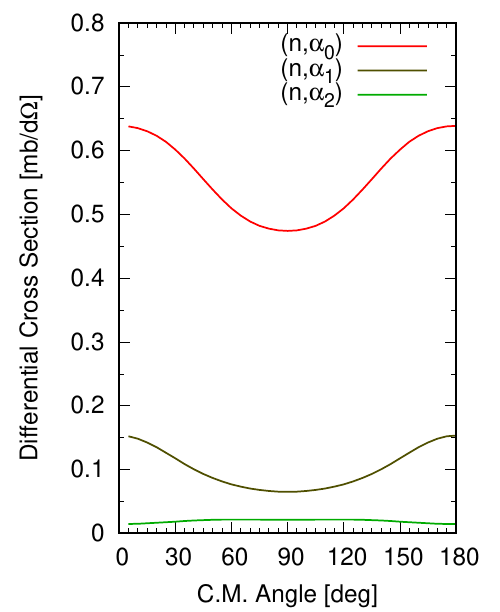}} \\
    \end{tabular}
  \end{center}
  \caption{Calculated secondary particle angular distributions for the
    neutron-induced reactions on $^{58}$Ni at $E_n = 3$~MeV; neutron
    (left), proton (center), and $\alpha$-particle (right). }
  \label{fig:Ni58angdist}
\end{figure}

Figure~\ref{fig:Ni58} demonstrates some default calculations of
neutron-induced reactions on $^{58}$Ni, comparing with the evaluated
data in ENDF/B-VII.1 and JENDL-4.0, as well as experimental data in
literature (for the sake of simplicity, we use the same symbol for all
available experimental data points.) These are relatively well behaved
cases, and we suppose the other HF codes provide similar
predictions. CoH$_3$ also produces the emitted particle angular
distributions, which are shown in Fig.~\ref{fig:Ni58angdist}. The left
panel shows the neutron elastic scattering that includes both the
shape and compound elastic scattering cross sections, and the
inelastic scattering to the first, second and third excited states of
$^{58}$Ni.  The center panel is for the proton and the right is the
$\alpha$-particle. The scattering angular distribution in a compound
reaction process $a+A \to b+B$ is calculated with the Blatt-Biedenharn
formalism~\cite{Kawano2015c},
\begin{equation}
  \left(\frac{d\sigma}{d\Omega}\right)_{ab}
   = \sum_L B_L P_L(\cos \theta_b) \ ,
 \label{eq:CNleg}
\end{equation}
The $B_L$ coefficient is given by Moldauer's statistical theory as
\begin{eqnarray}
B_L &=& \frac{1}{4k^2} \frac{(-)^{I_B - I_A + s_b - s_a}}{(2s_a+1)(2I_A+1)}
    \sum_J (2J+1)^2  \frac{1}{N_J} \nonumber\\
   &\times& \sum_{l_aj_a}  \sum_{l_bj_b}  W_{ab}
    \left\{
      X_{l_aj_a}(E_a) X_{l_bj_b}(E_b) +  \delta_{I_AI_B}\delta_{E_aE_b} Y_{l_aj_a,l_bj_b}(E_a,E_b)
    \right\} \ ,
\end{eqnarray}
where $k$ is the incident particle wave number, 
 $W_{ab}$ is the width fluctuation correction factor, $I$ and $s$ are 
the spin of nucleus and particles, and 
\begin{eqnarray}
 X_{lj}(E) &=& Z(ljlj;sL) W(jJjJ;IL) T_{lj}(E) \ , \\
 Y_{l_aj_a,l_bj_b}(E_a,E_b) &=& (1-\delta_{l_al_b})(1-\delta_{j_aj_b})
   \left\{Z(l_aj_al_bj_b;s_aL) W(Jj_aJj_b;I_AL)\right\}^2 \nonumber \\
 &\times& T_{l_aj_a}(E_a) T_{l_bj_b}(E_b)  \ ,
\end{eqnarray}
where $T_{lj}$ is the transmission coefficient, $Z$ is the
$Z$-coefficients, and the normalization $N_J$ is given by integrating
and summing all possible decay channels from the compound state $J$,
\begin{equation}
 N_J = \sum \int T_{lj}(E) dE .
\end{equation}
For the Hauser-Feshbach theory, $W_{ab}=1$ and
$Y_{l_aj_a,l_bj_b}(E_a,E_b)=0$.  In Fig.~\ref{fig:Ni58angdist} case,
the $\alpha$-particle emission that leaves the residual nucleus in its
ground state, the $(n,\alpha_0)$ reaction, shows large anisotropy
\cite{Kawano1999b}.

\section{Diagonalization of Coupled-Channels $S$-Matrix}

\begin{figure}[h]
  \begin{center}
    \begin{tabular}{cc}
      \resizebox{0.45\textwidth}{!}{\includegraphics{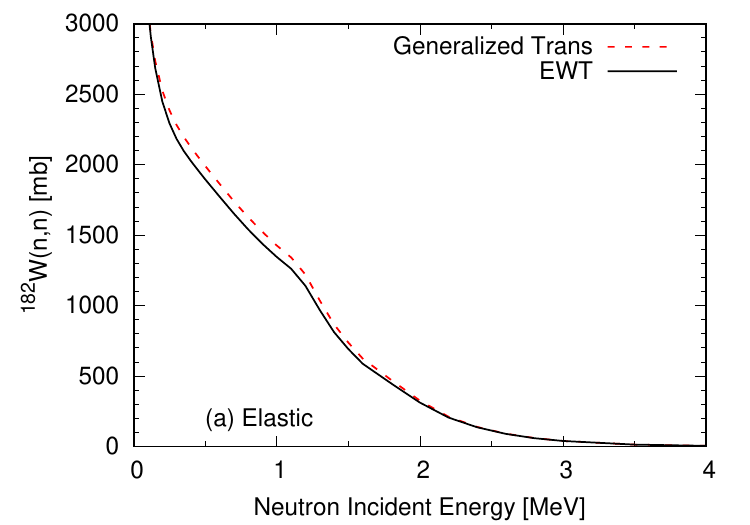}} &
      \resizebox{0.45\textwidth}{!}{\includegraphics{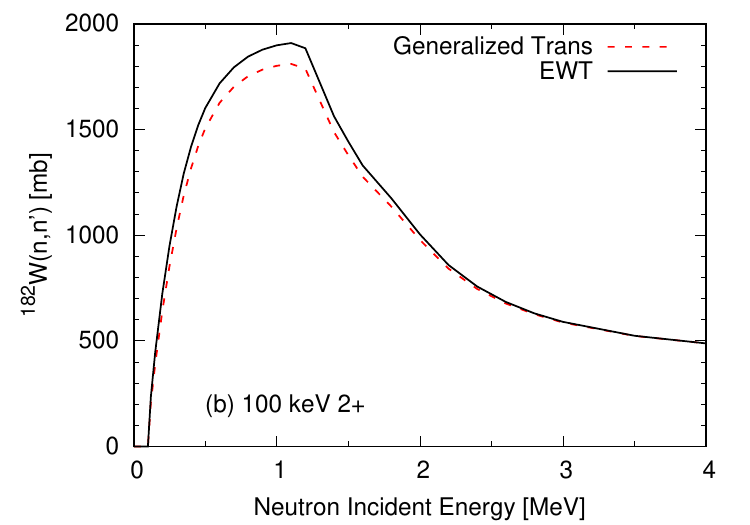}}\\
      \resizebox{0.45\textwidth}{!}{\includegraphics{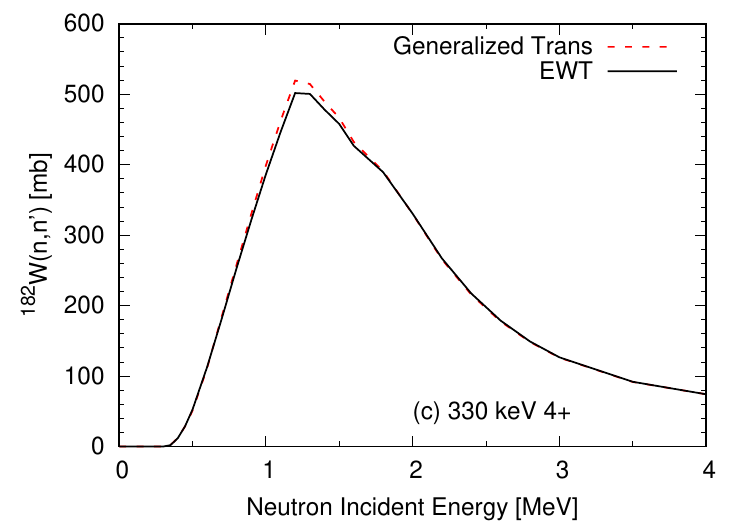}} &
      \resizebox{0.45\textwidth}{!}{\includegraphics{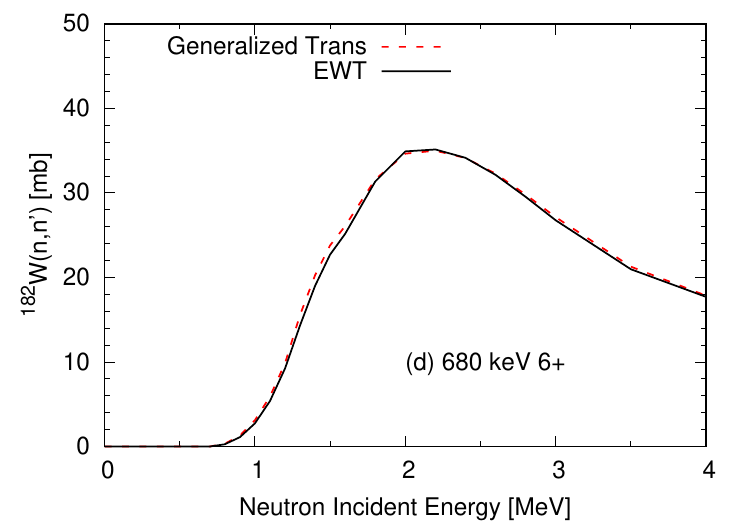}}\\
    \end{tabular}
  \end{center}
  \caption{Comparisons of the calculated (a) elastic and (b) -- (d)
    inelastic scattering cross sections for $^{182}$W. The solid
    curves are the full Engelbrecht-Weidenm\"{u}ller transformation
    (EWT) case, while the dashed curves are for the generalized
    transmission coefficient case.}
  \label{fig:w182inel}
\end{figure}

When strongly coupled channels exist, such as the direct inelastic 
scattering to the collective states, the scattering $S$-matrix
contains some off-diagonal elements, hence we cannot apply the standard
HF formalism.  In CoH$_3$, the coupled-channels $S$-matrix
is transferred into the diagonalized eigen-channel space (EWT). 
Since Satchler's penetration matrix
\begin{equation}
  P_{ab} = \delta_{ab} - \sum_c \ave{S_{ac}}\ave{S_{bc}^*} \ ,
\end{equation}
is Hermitian, this can be diagonalized by a unitary
transformation~\cite{Engelbrecht1973}
\begin{equation}
  (UPU^\dag)_{\alpha\beta} = \delta_{\alpha\beta} p_\alpha \ ,
  \qquad 0 \le p_\alpha \le 1 \ ,
\end{equation}
and the same matrix $U$ diagonalizes the scattering matrix,
\begin{equation}
  \ave{\tilde{S}} = U \ave{S} U^T \ .
\end{equation}
Here the Roman letters are for the channel index in the physical
space, and the Greek letters are for the eigen-channel. The width
fluctuation correction is performed in the eigen-channel, and they are
transformed back to the physical space
\begin{equation}
  \sigma_{ab}
  = \sum_{\alpha\beta\gamma\delta} U_{\alpha a}^* U_{\beta b}^* U_{\gamma a} U_{\delta b}
    \ave{ \tilde{S}_{\alpha\beta} \tilde{S}_{\gamma\delta}^* } \ ,
  \label{eq:backtrans1}
\end{equation}
where $\ave{ \tilde{S}_{\alpha\beta} \tilde{S}_{\gamma\delta}^* }$ is
the width fluctuation corrected cross section in the eigen-channel.
Rewriting Eq.~(\ref{eq:backtrans1}) into more convenient form includes
a term $\ave{\tilde{S}_{\alpha\alpha}\tilde{S}_{\beta\beta}^*}$, and
we estimated this average by applying the GOE
technique~\cite{Kawano2016}.

This transformation is still optional, since it requires longer
computational time when the number of coupled-channels is large.  When
the transformation is not activated, CoH$_3$ calculates the
generalized transmission coefficients from the coupled-channels
$S$-matrix, where the direct reaction components are eliminated from
the compound formation cross section \cite{Kawano2009}, and a usual HF
calculation is performed. This approximation works well when the
target nucleus is not so strongly deformed. Figure~\ref{fig:w182inel}
shows comparisons of the calculated elastic and inelastic scattering cross
sections for the strongly deformed $^{182}$W, and two cases are given;
the EWT case (solid curves) and the generalized transmission
coefficients (dashed curves).  A relatively large difference is seen
in the first excited state case.

\section{Conclusion}

We outlined the coupled-channels Hauser-Feshbach code, CoH$_3$. The
code includes several models that are indispensable for producing
evaluated nuclear data in the keV to tens of MeV region. The code is
designed to fully utilize the coupled-channels calculation, which is
especially important for evaluating nuclear data of deformed nuclei
such as actinides. As an example, calculations for the neutron-induced
elastic and inelastic scattering on $^{182}$W were shown, where 
two methods implemented in CoH$_3$ to combine the coupled-channels and
the Hauser-Feshbach theories are employed.

\section*{Acknowledgment}

This work was carried out under the auspices of the
National Nuclear Security Administration of the U.S. Department of
Energy at Los Alamos National Laboratory under Contract
No. 89233218CNA000001.


\end{document}